# Finite Volume Physical Informed Neural Network (FV-PINN) with Reduced Derivative Order for Incompressible Flows

Zijie Su [1], Yunpu Liu [1], Sheng Pan [3], Zheng Li [1,*] and Changyu Shen [1,2]

[1] State Key Laboratory of Structural Analysis, Optimization and CAE Software for Industrial Equipment, Department of Engineering Mechanics, Dalian University of Technology, Dalian 116024, China
[2] School of Materials Science and Engineering, The Key Laboratory of Material Processing and Mold of Ministry of Education, Zhengzhou University, Zhengzhou 450002, China
[3] Department of Micro Engineering, Kyoto University, 615-8540, Kyoto, Japan
* Correspondence: lizheng@dlut.edu.cn;

**Abstract:** Physics-Informed Neural Networks (PINN) has evolved into a powerful tool for solving partial differential equations, which has been applied to various fields such as energy, environment, engineering, etc. When utilizing PINN to solve partial differential equations, it is common to rely on Automatic Differentiation (AD) to compute the residuals of the governing equations. This can lead to certain precision losses, thus affecting the accuracy of the network prediction. This paper proposes a Finite Volume Physics-Informed Neural Network (FV-PINN), designed to address steady-state problems of incompressible flow. This method divides the solution domain into multiple grids. Instead of calculating the residuals of the Navier-Stokes equations at collocation points within the grid, as is common in traditional PINNs, this approach evaluates them at Gaussian integral points on the grid boundaries using Gauss's theorem. The loss function is constructed using the Gaussian integral method, and the differentiation order for velocity is reduced. To validate the effectiveness of this approach, we predict the velocity and pressure fields for two typical examples in fluid topology optimization. The results are compared with commercial software COMSOL, which indicates that FVI-PINN significantly improves the prediction accuracy of both the velocity and pressure fields while accelerating the training speed of the network.

## 1. Introduction

In recent years, Physics-Informed Neural Network (PINN) (1) has emerged as a promising numerical method that, unlike traditional data-driven machine learning techniques, directly incorporates governing equations and boundary conditions into the loss function. This integration enhances model interpretability and ensures that predictions are more consistent with physical laws. PINN can provide accurate solutions to partial differential equations with minimal or even no prior data. Compared to conventional discrete numerical methods, such as finite difference and finite element methods, PINNs offer greater flexibility in handling complex geometries, as they do not require mesh generation. Furthermore, when applied to high-dimensional problems, PINNs effectively overcome the curse of dimensionality, making them especially suited for tackling such challenges(2). Consequently, PINNs have found widespread applications in various fields, including diffusion equations, materials science, quantum mechanics, solid mechanics, and fluid dynamics(3–7) .

In particular, researchers have successfully applied PINNs to a wide range of fluid dynamics problems, demonstrating strong performance in both laminar and turbulent flow regimes. For example, Raissi et al. (1) pioneered the use of PINNs to model the Navier-Stokes equations, ,enabling the prediction of velocity and pressure fields for incompressible flows. Building on this, Rao et al. (8) proposed a mixed-variable scheme of PINNs for fluid dynamics, applying it to simulate steady and transient laminar flows at low Reynolds numbers. Jin et al. (9) introduced Navier-Stokes flow nets (NSFnets) to model high Reynolds number turbulence, improving prediction accuracy by dynamically adjusting the weights between data and physics. Cai et al. (10) developed a method using

PINNs to infer full continuous three-dimensional velocity and pressure fields from 3-D temperature field snapshots obtained through Tomographic Background Oriented Schlieren (Tomo-BOS) imaging. Additionally, Mao et al. (11) used PINNs to solve both forward and inverse high-speed aerodynamic flow problems by approximating the Euler equations.

However, most existing models based on PINNs for solving partial differential equations achieve this by directly embedding the residuals of the governing equations at the collocation points into the loss function(8,12). When applying PINNs to solve incompressible flow, the mass conservation equation Eq.(1) is typically enforced by treating the stream function $\phi$ of the velocity as the network output. By differentiating $\phi$ with respect to the spatial coordinates, the velocity components in each direction can be obtained (3). To compute the residuals of the Navier-Stokes equations, the second derivatives of the velocity must be calculated, which effectively requires computing the third derivatives of the network outputs. In this context, Automatic Differentiation (AD) is commonly used to compute the derivatives of the network output with respect to the network inputs(13). To compute first-order derivatives, AD uses both forward and backward passes. In solving high-order partial differential equations, AD can be recursively applied $n$ times to compute $n$th-order derivatives, leading to substantial computational overhead and potential accuracy loss(14). To address this issue, numerous researchers (15–19) have proposed loss functions in weak form, which reduce the order of derivatives required for the PDEs. Several examples have demonstrated the potential of this approach in solving high-dimensional partial differential equations.

Inspired by the finite volume method, we propose the FV-PINN. This method leverages Gauss's theorem to reduce the order of the Navier-Stokes equations and employs the reduced integral expressions as the loss function for PINN, thereby enabling unsupervised training. Without requiring any prior data, the steady-state velocity and pressure fields of incompressible laminar flow can be predicted solely based on the reduced integral expressions of the Navier-Stokes equations and the problem's boundary conditions. To systematically explain this theory and verify its effectiveness, Chapter 2 introduces the theoretical framework, loss function construction, and sampling strategies of FV-PINN. Chapter 3 demonstrates the reliability and effectiveness of the FV-PINN model through two numerical examples. Finally, the main contributions and findings of this study are summarized in the concluding chapter, along with a discussion of the potential of this method to be further extended to fluid topology optimization.

**2. Finite Volume Physical-Informed Neural Network (FV-PINN)**

*2.1 Governing equation*

In this paper, we present FV-PINN, which is used to predict the velocity and pressure fields for incompressible laminar flow under steady-state conditions. Let us first consider the governing equations of this problem in the steady-state case:

$$\nabla \cdot \mathbf{u} = 0 \tag{1}$$

$$(\mathbf{u} \cdot \nabla)\mathbf{u} = -\frac{1}{\rho}\nabla p + \frac{\mu}{\rho}\nabla^2 \mathbf{u} \tag{2}$$

Where $\nabla$ is the Nabla operator, $\mathbf{u} = (u_1, u_2)$ is the velocity vector (in 2D problem), $p$ is the pressure, $\mu$ is the viscosity of the fluid, $\rho$ is the density of fluid. To naturally satisfy the mass conservation equation (1), researchers(1,8) define the stream function $\phi$ of the velocity as the output of the neural network. This allows the velocity components in each direction to be obtained through Eq. (3).

$$u_1, u_2 = \frac{\partial \phi}{\partial y}, -\frac{\partial \phi}{\partial x} \tag{3}$$

*2. 2 Integral Formulation of the Navier-Stokes Equations Using the Finite Volume Method*

In solving the Navier-Stokes equations, the Finite Volume Method applies integration of the governing equations over discrete control volumes, converting the differential form into an integral form. This transformation facilitates the handling of complex boundaries and discontinuities. For incompressible flows, to satisfy the mass conservation equation, the stream function $\phi$ is typically used as the output in PINNs, as previously discussed, eliminating the need for further adjustments. As for the momentum conservation equation, the divergence theorem(20), also known as Gauss's theorem, is applied as shown below to convert volume integrals into boundary integrals, which correspond to the flux of the momentum conservation equations on the cell boundaries.

$$\int_V (\nabla \cdot \mathbf{u})dV = \oint_S \mathbf{u} \cdot \mathbf{n}dS \qquad (4)$$

The following section addresses each term in the momentum equation Eq. (2). The inertial term $(\mathbf{u} \cdot \nabla)\mathbf{u}$ from momentum equation Eq. (2) is transformed into a surface flux form through Gauss's theorem:

$$\int_{\Omega_i} (\mathbf{u} \cdot \nabla)\mathbf{u} dV = \oint_{\partial\Omega_i} \mathbf{u}(\mathbf{u} \cdot d\mathbf{A}) \qquad (5)$$

The pressure gradient $\nabla p$ is also converted into a surface flux:

$$\int_{\Omega_i} \frac{1}{\rho}\nabla p dV = \oint_{\partial\Omega_i} \frac{p}{\rho} d\mathbf{A} \qquad (6)$$

For the viscous diffusion term, it is treated as a surface flux on the boundary:

$$\int_{\Omega_i} \frac{\mu}{\rho}\nabla^2 \mathbf{u} dV = \oint_{\partial\Omega_i} \frac{\mu}{\rho}\nabla \mathbf{u} \cdot d\mathbf{A} \qquad (7)$$

After these transformations, the momentum conservation equation becomes an integral equation over the control volume, expressed the fluxes on the control volume boundary.

$$\int_\Omega \left[(\mathbf{u} \cdot \nabla)\mathbf{u} + \frac{1}{\rho}\nabla p - \frac{\mu}{\rho}\nabla^2 \mathbf{u}\right] dV = \oint_{\partial\Omega} \mathbf{u}(\mathbf{u} \cdot d\mathbf{A}) + \oint_{\partial\Omega} \frac{p}{\rho} d\mathbf{A} - \oint_{\partial\Omega} \frac{\mu}{\rho}\nabla \mathbf{u} \cdot d\mathbf{A} \qquad (8)$$

According to the above modifications, the two-dimensional form of the boundary flux integral is as follows:

$$\int_\Gamma \left[u_1(u_1 n_1 + u_2 n_2) - \frac{\mu}{\rho}\left(\frac{\partial u_1}{\partial x_1} n_1 + \frac{\partial u_1}{\partial x_2} n_2\right) + p n_1\right] d\Gamma \quad x\ direction \qquad (9)$$

$$\int_\Gamma \left[u_2(u_1 n_1 + u_2 n_2) - \frac{\mu}{\rho}\left(\frac{\partial u_2}{\partial x_1} n_1 + \frac{\partial u_2}{\partial x_2} n_2\right) + p n_2\right] d\Gamma \quad y\ direction \qquad (10)$$

Here, $u_1$ and $u_2$ represent the velocities in the $x$- and $y$- directions at each integration point, respectively; $n_1$ and $n_2$ are the components of the normal vector to the cell boundary in the $x$- and $y$-direction; and $p$ is the pressure at this point.

## 2.3 Implementation of FV-PINN

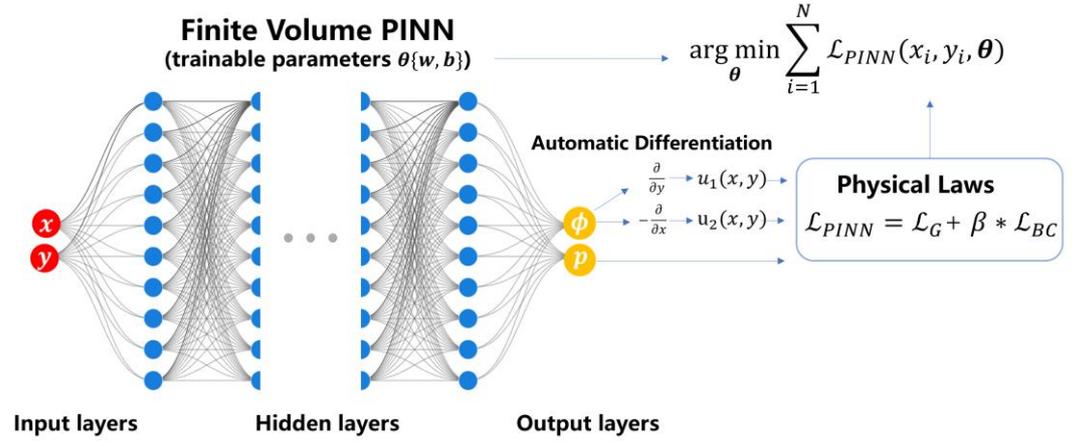

**Figure 1 Schematic of FV-PINN.** Here, $x$ and $y$ are the horizontal and vertical coordinates of the flow field integration points. $\boldsymbol{\theta}\{w, b\}$ represents the weights and biases of FV-PINN, which are continuously updated during training.

In the present work, the proposed FV-PINN takes a standard multilayer fully connected neural network as a prototype (shown in Figure 1). In this setup, the neural network consists of multiple hidden layers, where each layer is fully connected to the next. Mathematically, the operation of each hidden layer can be represented as follows:

$$\boldsymbol{h}^{(l+1)} = \sigma(\boldsymbol{W}^l \boldsymbol{h}^l + \boldsymbol{b}^l) \tag{11}$$

Where $\boldsymbol{h}^{(l+1)}$ denotes the activations in the $l$-th layer, $\boldsymbol{W}^l$ and $\boldsymbol{b}^l$ are the weight matrix and bias vector for that layer, and $\sigma$ is the activation function applied element-wise. For this network, we use the hyperbolic tangent (tanh) as the activation function which is suitable to capture nonlinear patterns in flow field, as shown below:

$$\sigma(x) = \tanh(x) = \frac{e^x - e^{-x}}{e^x + e^{-x}} \tag{12}$$

In the context of solving PDEs with FV-PINN, we use the Mean Squared Error (MSE) loss to quantify the residuals of the governing equations. The MSE loss is calculated as:

$$MSE = \frac{1}{N} \sum_{i=1}^{N} (\mathcal{R}_i)^2 \tag{13}$$

where $\mathcal{R}_i$ represents the contributions of convective momentum flux, pressure flux, and viscous momentum flux across the boundaries of the $N$-th cell. And $N$ is the number of elements.

The network input is the spatial coordinates $\boldsymbol{x} = (x, y)$ of the sampled points in the flow field, and the outputs are the stream function $\phi$ and pressure $p$, as formulated in Eq.(1) and Eq.(2). This approach allows us to obtain a continuous solution that satisfies the weak form of the NS equations. To compute the derivatives of the output quantities with respect to the spatial coordinates, we leverage automatic differentiation, which efficiently calculates both first- and second-order derivatives. Specifically, given the neural network output $f(\boldsymbol{x}, \boldsymbol{\theta})$, where $\boldsymbol{x}$ represents the spatial coordinates and $\boldsymbol{\theta}$ denotes the network parameters, automatic differentiation provides the derivatives:

$$\frac{\partial f}{\partial x}, \quad \frac{\partial^2 f}{\partial x^2}, \quad \frac{\partial^2 f}{\partial x \partial y}, \quad \text{and so on.} \tag{14}$$

These derivatives are then substituted into the integral expressions of the finite volume formulations shown in Eq.(9) and Eq.(10), allowing us to evaluate the governing equations over each cell. By doing so, we ensure that the network output satisfies the integral form of the conservation laws within each finite volume cell.

Furthermore, we introduce a boundary condition loss function $\mathcal{L}_{BC}$, which, along with the physical loss $\mathcal{L}_G$, constitutes the total loss function $\mathcal{L}_{PINN}$:

$$\mathcal{L}_{PINN} = \mathcal{L}_G + \beta * \mathcal{L}_{BC} \tag{15}$$

The physical loss $\mathcal{L}_G$ is computed by evaluating the Navier-Stokes equation residuals after transforming them via the divergence theorem into integrals, which are then approximated using Gaussian quadrature. The boundary condition loss function $\mathcal{L}_{BC}$ is computed by substituting the velocity field and pressure field predicted by the FV-PINN into either Dirichlet or Neumann boundary conditions, depending on the specific problem. β is a constant that balances the contributions of the two types of loss functions.

The PINN minimizes the loss function through gradient descent algorithms, such as Adam and L-BFGS (21,22). Through iterative optimization of the network parameters, the predicted solution gradually converges towards the true physical solution.

*2.4 Gaussian Quadrature-based Sampling Strategy and Loss Function Computation*

Unlike the uniform sampling of points in the entire flow field in PINN for fluid(1), in this study, the selection of sampling points is based on Gaussian Quadrature(23). Specifically, like Finite Volume Method, the flow field is discretized into elements, and the sampling points involved in the FV-PINN loss function calculation are located on the boundaries of these elements, as shown in Figure 2. These points are selected through Gaussian Quadrature to ensure that the approximation of physical quantities at each integration point achieves higher accuracy. Gaussian Quadrature allows for efficient sampling in high-dimensional spaces, especially when dealing with complex boundary conditions and flow fields, significantly reducing errors.

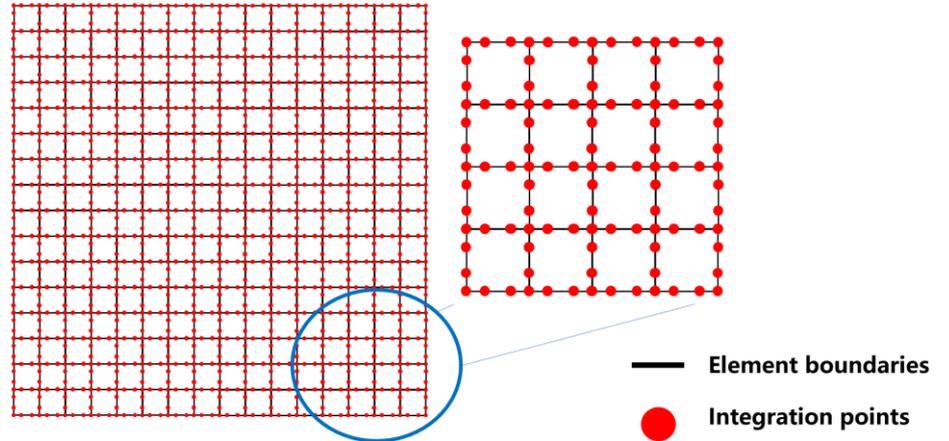

**Figure 2** Diagram of integration points.

Through Gaussian Quadrature, we can convert the loss function from an integral form into a discrete numerical form. By applying the Gaussian Quadrature coefficients at the sampling points, we can compute the contribution of loss function at each point, and thus obtain the loss function of the governing physical equations, as shown in Eq.(16).

$$\begin{aligned}
\mathcal{L}_G &= \sum_{j=1}^{N} \int_{\partial V_i} (I_1 + I_2) dS \\
&= \sum_{j=1}^{N} \int_{\partial V_j} \left[ \left( u_1(u_1 n_1 + u_2 n_2) - v\left(\frac{\partial u_1}{\partial x_1} n_1 + \frac{\partial u_1}{\partial x_2} n_2\right) + p n_1 \right) \right. \\
&\quad + \left. \left( u_2(u_1 n_1 + u_2 n_2) - v\left(\frac{\partial u_2}{\partial x_1} n_1 + \frac{\partial u_2}{\partial x_2} n_2\right) + p n_2 \right) \right] dS \\
&\approx \sum_{j=1}^{N} \sum_{k=1}^{n} \left[ \left( u_1(u_1 n_1 + u_2 n_2) - v\left(\frac{\partial u_1}{\partial x_1} n_1 + \frac{\partial u_1}{\partial x_2} n_2\right) + p n_1 \right) \right. \\
&\quad + \left. \left( u_1(u_1 n_1 + u_2 n_2) - v\left(\frac{\partial u_1}{\partial x_1} n_1 + \frac{\partial u_1}{\partial x_2} n_2\right) + p n_1 \right) \right]_{j,k} w_k
\end{aligned} \tag{16}$$

Where $N$ represents the number of elements in the flow field, $n$ denotes the number of integration points in each element, and $w_k$ represents the Gaussian quadrature weight.

## 3. Model Validation and Results Analysis

The Double Pipe and Pipe Bend are classic examples in fluid topology optimization, first introduced by Borrvall and Peterson in 2003(24). In the following chapter, we will use the proposed FV-PINN to predict the steady-state velocity and pressure fields for these two cases and compare the results with the commercial software COMSOL, thereby validating its effectiveness. The FV-PINN is implemented through PyTorch and trained on NVIDIA RTX 4090.

### 3.1 Pipe Bend Problem

This problem is illustrated in Figure 3.a. On the left-hand side, there is an inlet at which parabolic normal velocity profiles are prescribed with a maximum velocity of $U_{in}(U_{max} = 1\ m/s)$. On the bottom, there is a zero-pressure outlet at which the flow is specified to exit in the normal direction. For the steady case, the dynamic viscosity and density of the fluid is $0.02\ kg/(m \cdot s)$ and $1\ kg/m^3$ respectively.

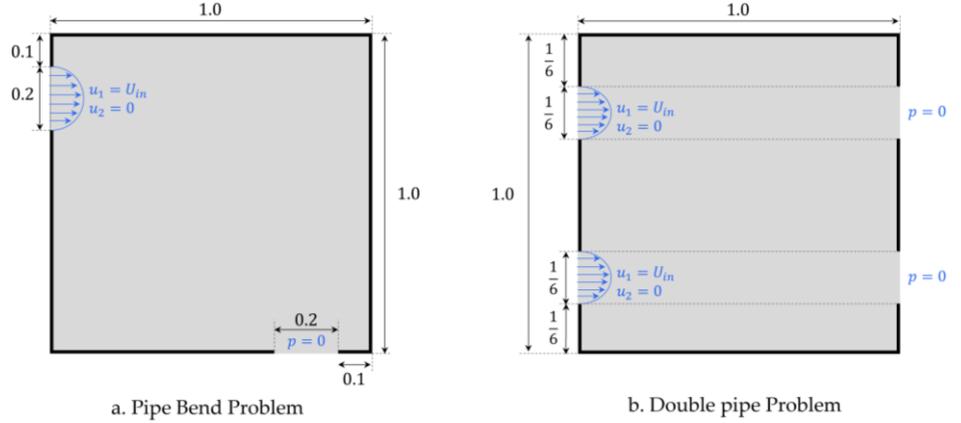

**Figure 3** Problem setup

For the boundary conditions, instead of performing additional sampling, we select integration points located at the boundaries, compute the associated boundary condition loss terms $\mathcal{L}_{BC}$, and incorporate them into the final loss function. The Adam optimizer is used for training. The specific neural network parameters are as follows:

**Table 1. The parameters of FVI-PINN**

| Parameter | Value |
| --- | --- |
| Architecture: layers and neurons | [2,40,40,40,40, 40,40,40,40,2] |
| Training epochs | 20000 |
| Learning rate | 0.003 |
| $\beta$ (scaling $\mathcal{L}_G$ and $\mathcal{L}_{BC}$) | 0.0001 |

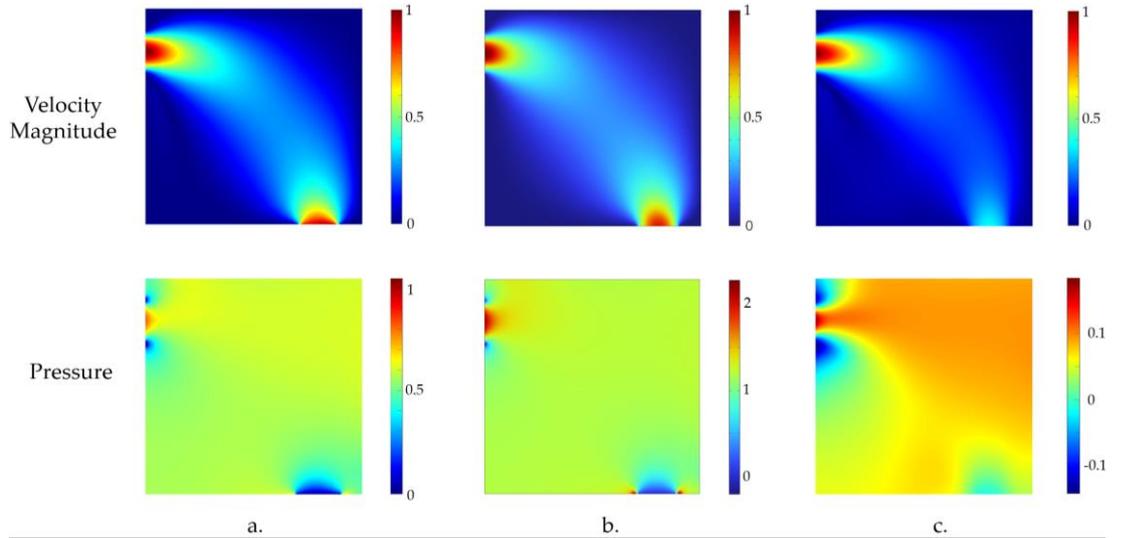

**Figure 4. Comparison of Results for the Pipe Bend Problem:** (a) Predicted results from FV-PINN, (b) Computational results from COMSOL, and (c) Predicted results from the traditional NS equation-based PINN.

Figure 4 shows the prediction results of FV-PINN for the Pipe Bend problem. As can be seen, accurate predictions of the velocity field are achieved, while the pressure field can be reasonably predicted qualitatively. The numerical predictions of the pressure field differ from those of COMSOL. This phenomenon has already been explained by Raissi in (1), where it was stated that the absolute pressure field in incompressible flows is indeterminate and can only be predicted up to an arbitrary constant. The model predicts the relative pressure variations, but the absolute pressure value remains undetermined. It can be also observed that the traditional PINN, which directly incorporates the residuals of the Navier–Stokes equations into the loss function, fails to provide accurate predictions for both the velocity and pressure fields in this problem. Consequently, it can be concluded that FV-PINN, by reducing the order of differentiation in automatic differentiation, significantly enhances the accuracy of predictions for incompressible flow fields.

Table 2 presents a comparison between FV-PINN and the traditional NS-based PINN in terms of training time and the number of sampling points (with their respective prediction results shown in Figure 4.a and Figure 4.c). As shown in the table, after 20,000 iterations, FV-PINN achieves more accurate predictions than the traditional NS-based PINN while utilizing fewer sampling points, resulting in a 40% reduction in training time. Figure 5 illustrates the variations of the loss functions during the training process for both models, demonstrating that FV-PINN exhibits significantly faster convergence compared to the traditional NS-based PINN.

**Table 2. Comparison of Training Time and Sampling Points**

| Parameter | Training Time(s) | The number of Sampling Points |
|---|---|---|
| FV-PINN | 442.72 | 29280 |
| Traditional PINN | 745.77 | 42202 |

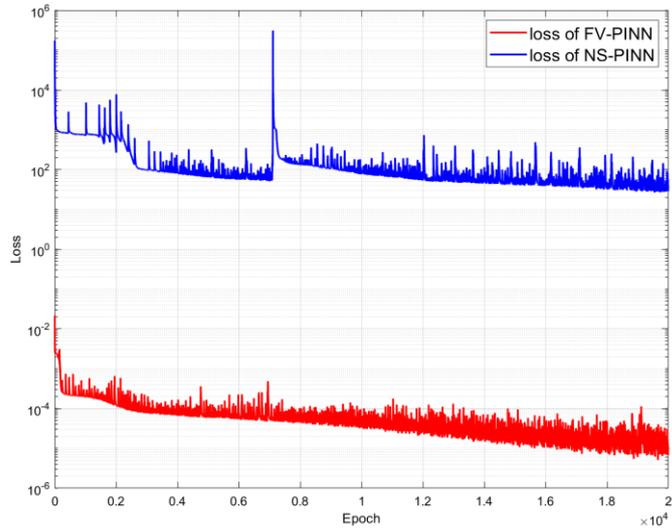

**Figure 5.** Comparison of Loss Functions: FV-PINN vs. Traditional PINN

*3.2 Double Pipe Problem*

This problem is illustrated in Figure 3.b. In this subsection, the Double Pipe Case is revisited to further substantiate the conclusions established in the previous section. This configuration features two inlets, where the velocity magnitude and distribution are consistent with those in the Pipe Bend Case, while the outlets are subject to zero-pressure boundary conditions. All other physical parameters, as well as the neural network architecture and hyperparameters, are identical to those in the preceding subsection.

Figure 6 presents a comparative analysis of the FV-PINN predictions, the results obtained using the commercial FVM-based software COMSOL, and the computations based on the Finite Element Method (FEM) performed in MATLAB. The findings demonstrate that FV-PINN achieves outstanding predictive performance in the Double Pipe Case. Notably, in terms of pressure field predictions, FV-PINN exhibits greater consistency with COMSOL results compared to the FEM.

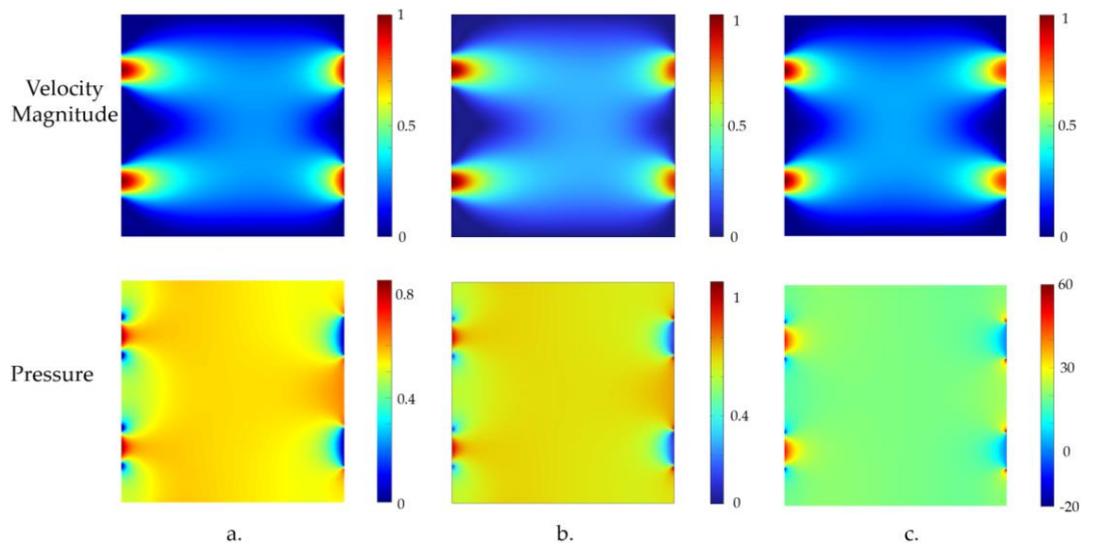

**Figure 6 Comparison of Results for the Double Pipe Problem:** (a) Predicted results from FV-PINN, (b) Computational results from COMSOL, and (c) Results obtained using the Finite Element Method (FEM) in MATLAB(25).

*3.3 Summary of Case Study Results*

The predictions of the velocity and pressure fields obtained from FV-PINN model closely resemble those from COMSOL, demonstrating a high level of agreement. The accuracy of these predictions highlights the effectiveness of the PINN approach in capturing the underlying flow dynamics, making it comparable to traditional computational methods. Through comparison, it is evident that FV-PINN significantly outperforms traditional NS-based PINNs in terms of prediction accuracy, achieving more precise results with fewer computational resources. In addition to its enhanced accuracy, FV-PINN demonstrates a substantial reduction in time cost, as it requires fewer sampling points. Furthermore, FV-PINN exhibits a much faster convergence rate than conventional NS-based PINNs, making it a highly efficient approach for solving complex flow problems.

## 4. Conclusions

This paper presents FV-PINN, a novel Physics-Informed Neural Network inspired by the Finite Volume Method. By leveraging the divergence theorem to reformulate the residuals of Navier-Stokes equations, FV-PINN reduces the reliance on high-order derivatives and achieves enhanced prediction accuracy and faster convergence. The proposed method has been validated on steady-state incompressible laminar flow problems by comparing with commercial software COMSOL, demonstrating superior accuracy and efficiency compared to traditional PINNs. FV-PINN offers a promising framework for solving complex fluid dynamics problems with reduced computational cost and higher precision, paving the way for broader applications in computational physics.

In future work, we aim to leverage FV-PINN to achieve highly accurate predictions of velocity and pressure fields in flow scenarios with dynamically evolving fluid-solid interfaces. This will lay the foundation for fully AI-driven fluid topology optimization, offering a novel technical framework and solutions to the challenges of fluid mechanics design.